\documentclass[aip,reprint,graphicx]{revtex4-1}
\usepackage{amsmath} 
\usepackage{amssymb} 
\usepackage{graphicx} 
\usepackage{epsfig}
\usepackage{color}
\usepackage{epstopdf}
\usepackage{rotating}
\usepackage{bm}

\draft % marks overfull lines with a black rule on the right

\begin{document}

%\preprint{}

\title{Spin molecular-orbit coupling and magnetic properties of the decorated honeycomb layers of Mo$_3$S$_7$(dmit)$_3$ crystals} %Title of paper



\author{Jaime Merino*}
\email[*]{jaime.merino@uam.es}
%\thanks{}
\affiliation{Departamento de F\'isica Te\'orica de la Materia Condensada, Condensed Matter Physics Center (IFIMAC) and
Instituto Nicol\'as Cabrera, Universidad Aut\'onoma de Madrid, Madrid 28049, Spain}
\author{Anthony C. Jacko}
\affiliation{School of Mathematics and Physics, The University of Queensland, Brisbane, Queensland 4072, Australia}
\author{Amie L. Kohsla}
\affiliation{School of Mathematics and Physics, The University of Queensland, Brisbane, Queensland 4072, Australia}
\author{Arnaud Ralko} 
\affiliation{Institut N\'eel, UPR2940, Universit\'e Grenoble Alpes et CNRS, Grenoble, FR-38042 France}
\author{Ben J. Powell}
\affiliation{School of Mathematics and Physics, The University of Queensland, Brisbane, Queensland 4072, Australia}

\date{\today}

\begin{abstract}
We explore the magnetic properties of isolated $a-b$ planes of trinuclear organometallic crystals, Mo$_3$S$_7$(dmit)$_3$,  
in which an interplay of strong electronic correlations and spin molecular-orbital coupling (SMOC) occurs. The magnetic properties
can be described by a XXZ+120$^0$, $S=1$ Heisenberg model on a honeycomb lattice with single-spin anisotropy, $D$, 
which depends strongly on SMOC.
%We find from Schwinger boson mean-field theory, that a transition to a large-$D$ phase 
%occurs when $D>3J$, where $J$ is the nearest-neighbor spin exchange coupling. 
Based on {\it ab initio} estimates of SMOC 
in Mo$_3$S$_7$(dmit)$_3$ crystals, we predict 
that the honeycomb layers of Mo$_3$S$_7$(dmit)$_3$ are N\'eel ordered. However, in materials with a greater degree of 
magnetic frustration, N\'eel order can give way to a large-$D$ phase. 
\end{abstract}


\maketitle %\maketitle must follow title, authors, abstract and \pacs

\section{Introduction}

Strong spin orbit coupling (SOC) in weakly interacting systems can lead to unconventional 
insulating states such as topological insulators. Strongly correlated phases
such as the topological Mott insulator can emerge from the interplay of 
strong Coulomb repulsion and SOC which may be realized in Ir-based transition 
metal oxides.\cite{pesin2010} For instance, in Na$_2$IrO$_3$ and Li$_2$IrO$_3$ materials, 
SOC removes the orbital degeneracy of $5d$ electrons leading to $S=1/2$ pseudospins 
which interact through anisotropic and quantum compass exchange interactions on a 
honeycomb lattice. These materials are potential realizations 
of the Heisenberg-Kitaev model, closely related to the Kitaev model 
which sustains a spin liquid state\cite{kitaev}.

\begin{figure}
 \includegraphics[width=6cm]{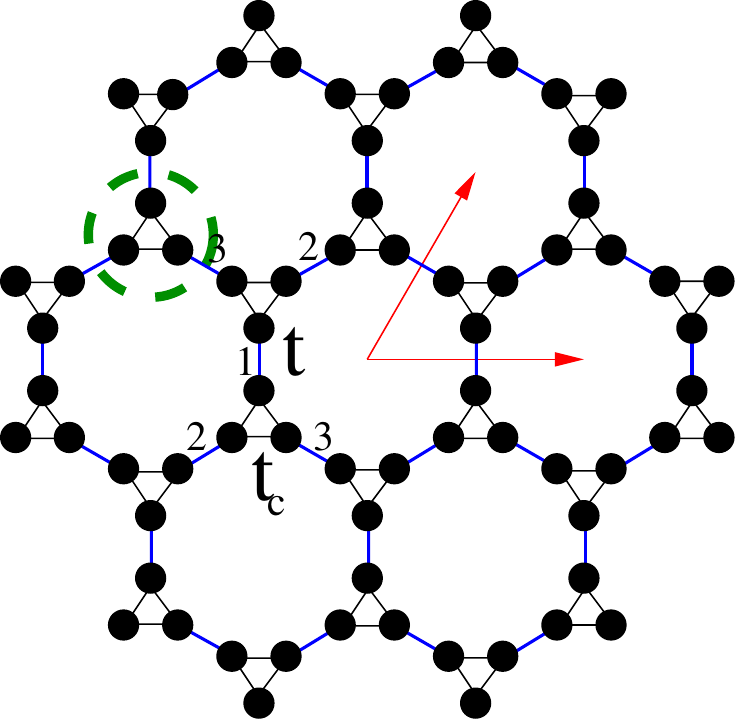}%
 \caption{The decorated honeycomb lattice realized in the $a$-$b$ planes of Mo$_3$S$_7$(dmit)$_3$ 
 crystals. Mo$_3$S$_7$(dmit)$_3$ molecules are described by three Wannier orbitals (black dots), which 
form  the triangles of the decorated lattice encircled by a dashed green line. The intracluster hopping amplitude
between two molecular orbitals inside the triangle is denoted by $t_c$ while the intercluster 
hopping amplitude between two orbital in neighboring molecules is denoted by $t$. The red arrows
indicate the Bravais lattice vectors of the underlying honeycomb lattice. The three 
bonds connecting equal sites between nearest-neighbor clusters are numbered from 1 to 3.
 \label{fig:decorated}}%
 \end{figure}

Multinuclear coordinated organometallic 
complexes are strongly correlated systems in which spin-orbit coupling (SOC) can be relevant. 
The multinuclear complex,  Mo$_3$S$_7$(dmit)$_3$, consists of
honeycomb networks of Mo$_3$S$_7$(dmit)$_3$ molecules stacked
on top of each other along the $c$-direction of the crystal. 
Since the Mo$_3$S$_7$(dmit)$_3$ molecules can be described by three Wannier
orbitals\cite{jacko2015}, their packing on the honeycomb lattices within
the layers lead to decorated honeycomb lattices, as shown in Fig. \ref{fig:decorated}.  
Based on perturbative expansions including Coulomb repulsion and SMOC, the effective spin exchange model which 
describes the magnetic properties of these layers is a $S=1$ XXZ+120$^0$ quantum compass model on the 
honeycomb lattice.\cite{merino2017,powell2017} 

In the present paper, we discuss the role played by the single-spin anisotropy induced by SMOC in the magnetic 
properties of the honeycomb layers of Mo$_3$S$_7$(dmit)$_3$ crystals shown in Fig. \ref{fig:decorated}.  
%We show that a transition from a N\'eel ordered phase 
 %to a large-$D$ phase occurs within these layers at a critical SMOC which is much larger than the SMOC present in the honeycomb layers of Mo$_3$S$_7$(dmit)$_3$.
 We predict that isolated honeycomb layers of Mo$_3$S$_7$(dmit)$_3$ crystals are N\'eel ordered but
increasing the magnetic frustration of the lattice can drive the system into a large-$D$ phase.

%\label{}

\section{Spin orbital molecular coupling in trinuclear organometallic complexes}

\subsection{Model for isolated trimers} 
The isolated triangular clusters can be described by a Hubbard-Heisenberg model in the presence of 
the spin molecular-orbit coupling (SMOC):
\begin{equation}
H=H_0+H_{SMOC}+H_{U-J_F}
\label{eq:ham}
\end{equation}
where:
\begin{equation}
H_0=-t_c \sum_{\langle ij \rangle \sigma} \left( a^\dagger_{i\sigma} a_{j\sigma}  + H.c. \right),
\end{equation}
with $t_c$ the hopping between hybrid metal-ligand orbitals at sites in the
triangular clusters  and $a^\dagger_{i\sigma}$ creates an electron at the $i$th Wannier 
orbital with spin $\sigma$.

An important question is how SOC affects the low energy states described by these Wannier orbitals. Generically the one-electron spin-orbit coupling takes the form $H_{SOC}=\sum_\alpha{\bm K}_\alpha\cdot{\bm s}_\alpha$, where ${\bm s}_\alpha $ is the  spin of the $\alpha ^\text{th}$ electron and $\bm K_\alpha$ is a  pseudovectorial operator that acts only on the spatial part of the wavefunction. For example, in the Pauli approximation,
\begin{equation}
\bm K=-{e \hbar \over 2 m^2 c^2 } [ {\bm p} \times {\bm \nabla} \phi({\bm r}) ],
\label{eq:socop}
\end{equation} 
where $\bm p$ is the single electron momentum operator and $\phi(\bm r)$ is the potential through which the electron moves.

Spherical symmetry implies that for an isolated atom ${\bm K}\propto{\bm L}$, the angular momentum operator. Thus, in {\it ab initio} calculations, it is common to write the SOC as a linear superposition of atomic contributions, for example, in the  mean-field Breit-Pauli approximation\cite{neese2005,veeman1970}  
\begin{equation}
H_{SOC}={e^2 \hbar \over 2 m^2 c^2 }\sum_A \sum_\alpha {Z^{eff}_{A} \over |\bm r_{\alpha}-{\bm r}_A|^3} {\bm l}_{A\alpha}\cdot {\bm s}_\alpha,
\label{eq:atsoc}
\end{equation}
where $Z^{eff}_A$ is the effective charge of nucleus $A$ which accounts for the screening effects of the rest of the electrons on the 
nuclear potential, ${\bm l}_{\alpha A}=({\bm r}_\alpha -{\bm R}_A)\times{\bm p}_\alpha $, is the angular momentum of the $\alpha ^\text{th}$ electron relative to the nucleus A at ${\bm R}_A$. %If one were to project onto  atomic orbitals, $|\psi^A_n\rangle$, and sum the contributions one would have
%\begin{equation}
%H_{SOC}=\sum_{An} \lambda_{An} \sum_\alpha {\bm l}_{A\alpha}\cdot {\bm s}_\alpha,
%\label{eq:atsoc}
%\end{equation}
%where  $\lambda_{An}={e^2 \hbar \over 2 m^2 c^2 }\langle \psi^{A}_{n} | {Z^{eff}_{A} \over |r_{i}-{\bm r}_A|^3} |\psi^A_n \rangle$ are the SOC constants.
%%Hence, this hamiltonian also neglects two-electron terms which are approximately taken into account 
%%by using screened nuclear charges, $Z^{eff}_A$. 
%%This mean-field approach has been used to describe SOC effects in molecules containing not too heavy elements.\cite{neese2005}
%However, Eq. (\ref{eq:atsoc}) neglects the effects of the potential arising from other ions in the molecule or solid. For electrons in extended molecular orbitals this approximation is a fraught with risk. More importantly, neither Eqs. (\ref{eq:atsoc1}) nor (\ref{eq:atsoc}) 
This atomic approach does not provide any insight into how SOC acts on the low-energy states described by the Wannier orbitals and requires detailed first principles calculations for each new material studied. A more 
chemically intuitive description should consider the coupling between the electronic spin and the molecular orbital degrees of freedom.

Symmetry  dictates the general form of the SMOC\cite{khosla2017}. For cyclic molecules with $C_N$ symmetry the coupling is\cite{khosla2017} 
\begin{eqnarray}
H_{\mathrm{SMOC}}
&=& 
\sum_{m=1}^{L}\sum_{\sigma=-1/2}^{1/2}\sigma\lambda^z_m\left(\hat c_{m\sigma}^\dagger \hat c_{m\sigma} - \hat c_{-m\sigma}^\dagger \hat c_{-m\sigma} \right)
\label{eq:Hodd}
\\&&
\notag\hspace*{-0.8cm}+
\frac{1}{2} \sum_{j=\frac12}^{L-\frac12}\left[\lambda_j^{\pm}\left(\hat c_{j+\frac12\downarrow}^\dagger \hat c_{j-\frac12\uparrow} 
+\hat c_{-j+\frac12\downarrow}^\dagger \hat c_{-j-\frac12\uparrow} \right)
+\textrm{H.c.}\right],
\end{eqnarray}
where $c^\dagger_{m\sigma}=i^{|m|}\sum_j a_{j\sigma}^\dagger e^{i2\pi j/N}/\sqrt{N}$.
For a C$_3$ symmetric molecule, such as Mo$_3$S$_7$(dmit)$_3$, this simplifies to
\begin{eqnarray}
H_{SMOC}
&=& \frac{\lambda_{xy}}{2} \left( \hat c_{1\downarrow}^\dagger \hat c_{0\uparrow} + \hat c_{0\downarrow}^\dagger \hat c_{-1\uparrow} 
+  \hat c_{0\uparrow} ^\dagger \hat c_{1\downarrow} + \hat c_{-1\uparrow} ^\dagger \hat c_{0\downarrow} \right) \notag\\&&
+\lambda_z \left(\hat c_{1\uparrow}^\dagger \hat c_{1\uparrow} - \hat c_{1\downarrow}^\dagger \hat c_{1\downarrow} - \hat c_{-1\uparrow}^\dagger \hat c_{-1\uparrow} + \hat c_{-1\downarrow}^\dagger \hat c_{-1\downarrow} \right),\notag\\
\label{eq:SMOC}
&\equiv& \sum_\alpha  \left[\lambda_{xy} \left(  L_\alpha ^x s_\alpha ^x + L_\alpha ^y s_\alpha ^y  \right)+ \lambda_z L_\alpha ^z s_\alpha ^z \right]
\end{eqnarray}
where  $L_\alpha ^{x,y,z}$ are the Cartesian components of the one-electron molecular orbital ``angular momentum'' operator for the $\alpha ^\text{th}$ electron, 
$\lambda_{xy}$, is the transverse SMOC and $\lambda_z$ the longitudinal SMOC. 

The SMOC hamiltonian (\ref{eq:SMOC})  gives a far clearer understanding of the relevant low-energy physics than the atomistic description (\ref{eq:atsoc}). This is because 
molecular orbitals, not atomic orbitals, are the natural basis for the discussion of molecular physics.
As  molecular orbitals for cyclic molecules are Bloch states of the Wanniers, $H_{SMOC}$  
describes the coupling of orbital currents running around the plane of the molecule to the electron's spin. The low-energy model constructed from Wannier spinors from  
four component relativistic band structure calculations\cite{jacko2017} of Mo$_3$S$_7$(dmit)$_3$ yields single molecule SMOC exactly as predicted by Eq. (\ref{eq:SMOC}). 
%The SMOC contribution (\ref{eq:SMOC}) could be inferred by projecting the SOC operator (\ref{eq:socop}) 
%onto the Wannier {\it orbital} basis using the multi-atom potential: $\phi({\bm r})=\sum_{A=1}^3 {Z^{eff}_A \over |{\bm r}-{\bm R}_A|}$ describing the
%triangular molecule which is different from Eq. (\ref{eq:atsoc}) which is the sum of atomic SOC. In particular, the 
%expectation values of electron momentum ${\bm p}$ will be very different from those for electrons orbiting a single
%atom, particularly for electrons near the Fermi energy. 
%Note that the form of $H_{SMOC}$ obtained 
%reflects an intrinsic property of the molecular structure of Mo$_3$S$_7$(dmit)$_3$ clusters since Eq. (\ref{eq:SMOC})  
%describes the coupling of orbital currents running around the plane of the molecule with the electron's spin. 
%Although the origin of SMOC in the clusters is similar to atomic SOC, the particular planar structure
%of molecules forming multinuclear organometallic compounds naturally leads to the form of the interaction
%$H_{SMOC}$ in Eq. (\ref{eq:SMOC}) considered here.

Finally, the Hubbard-Heisenberg contribution describing the Coulomb and exchange energy in each 
triangular cluster is:
\begin{equation}
H_{U-J_F}=U\sum_i n_{i\uparrow}n_{i\downarrow}+J_F\sum_{\langle ij \rangle} \left( {\bm S}_i\cdot {\bm S}_j-\frac{n_i n_j}{4} \right),
\end{equation}
where $U$ is the onsite Hubbard interaction, $J_F$ is an intracluster exchange interaction, 
and $n_{i\sigma}=a^\dagger_{i\sigma}a_{i\sigma}$ the number operator. The {\it direct} ferromagnetic exchange, $J_F<0$,
plays a crucial role in generating magnetic anisotropies.\cite{merino2017}
%It plays a similar role as the Hunds coupling 
%in transition metal oxides\cite{jackeli2009}, which also generates magnetic exchange anisotropies between spins in the lattice.

%Thus, in a molecule $V({\bm r})$ is simply a linear
%superposition of the atomic potentials, $V({\bm r})= {e^2 Z_K(r_{iK}) \over r_{iK}^3}}$, where $K$ denotes 
%a particular atomic nucleus and ${\bm r}_{iK}={\bm r}_i-{\bm R}_K$ is the relative distance between the
%electron position and the nucleus. :
%\begin{equation}
%H_{SOC}=\sum_{ K} \lambda_K \sum_{i=1}^3 {\bm l}_{Ki} {\bm s}_i
%\end{equation}

The nearest-neighbor triangular clusters are connected through the hopping amplitude, $t$:
\begin{equation}
H_{kin}=-t \sum_{\langle l m\rangle}  \sum_i \left( a^\dagger_{\ell i\sigma}a_{m i\sigma}+a^\dagger_{m i\sigma}a_{\ell i\sigma} \right),
\label{eq:dumb}
\end{equation}
where $\langle l m \rangle $ denotes two nearest-neighbor triangular clusters and $i$ a cluster site in a given molecule.

\subsection{Effective spin exchange model}
\begin{figure}
 \includegraphics[width=6cm]{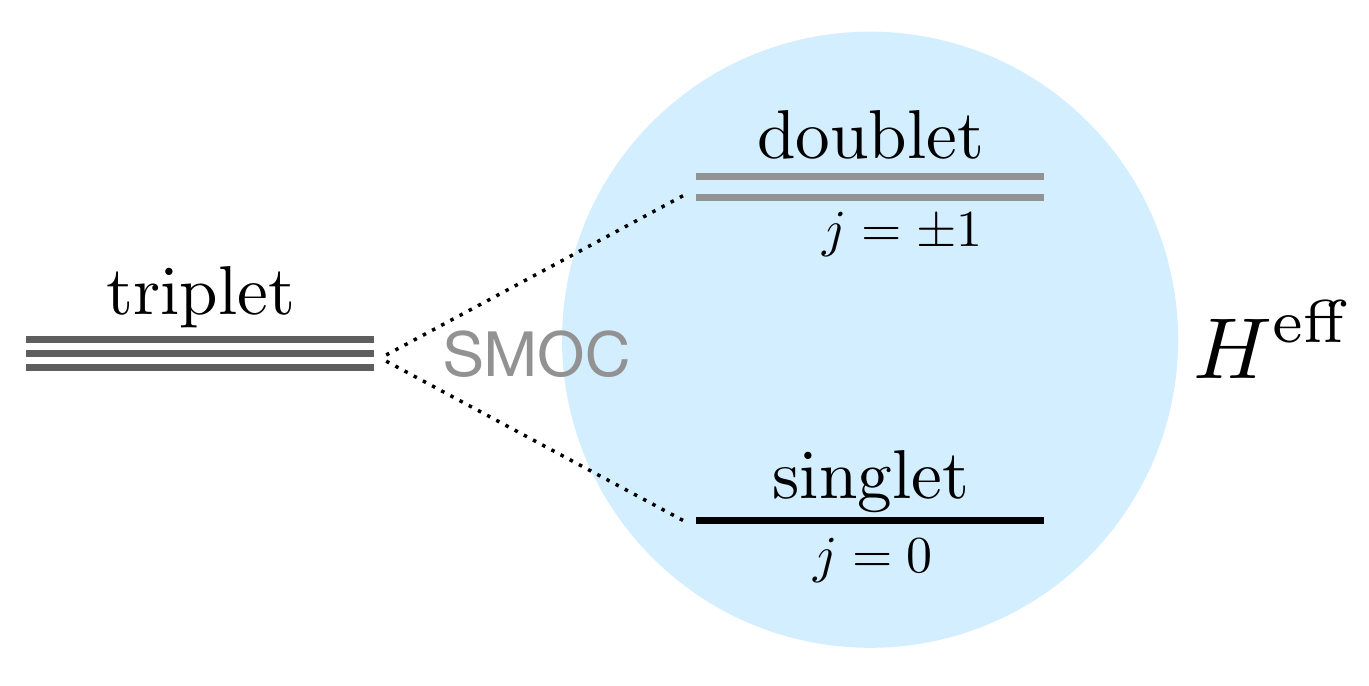}%
 \caption{\label{fig:levels} Effective low energy level spectrum of isolated trimers of Mo$_3$S$_7$(dmit)$_3$ molecules in the presence of SMOC. 
 %The ground state triplet of model (\ref{eq:ham}) is 
 %splitted by SMOC leading to a ground state singlet with pseudospin $j=0$ and a doublet with $j=\pm 1$ which can be described
% by the spin model (\ref{eq:trig}) with the single-spin anisotropy $D$. 
}%
 \end{figure}

By straightforward diagonalization of the full hamiltonian (\ref{eq:ham}), we have found that
in the strong coupling limit, $U \gg t$, isolated triangular clusters with four electrons effectively behave as pseudospin-1 localized
moments\cite{merino2017}. The many-body states of the cluster can be classified according to the $z$-component of the 
total angular momentum, $J_z=L_z+S_z$. The ground state of the isolated cluster with no SMOC is a triplet. As shown in Fig.\ref{fig:levels}, SMOC
splits the lowest energy triplet into a non-degenerate singlet ($j=0$) which is the ground state and a doublet ($j =\pm 1 $), 
where $j$ denotes the eigenstates of $J_z$. Since we have an even number of electrons in the 
cluster, Kramers theorem does not apply and non-degenerate states are possible. 
SMOC can be effectively accounted for as a single-spin anisotropy contribution at each cluster.
Hence, the effective spin model for the $m$-th Mo$_3$S$_7$(dmit)$_3$ molecule
in the crystal is just:
\begin{equation}
H_m^\text{eff}= D ({{\mathcal{ S}}}_{\bm r_m}^z)^2,
\label{eq:trig}
\end{equation} 
where ${{\mathcal{ S}}}_{\bm r_m}$ describes the effective pseudospin-1 localized at a triangular cluster.  
The single-spin anisotropy, $D$, depends strongly on SMOC as shown in Fig. \ref{fig:soc}.
Using perturbation theory \cite{powell2017,merino2017} we have obtained an effective spin exchange model describing the magnetic exchange coupling between the 
pseudospin-1 in the $a-b$ planes of Mo$_3$S$_7$(dmit)$_3$. This model is an XXZ+120$^0$ honeycomb quantum compass model \cite{merino2017}  
leading to:
\begin{widetext}
		\begin{eqnarray}
	H&=&  \sum_\ell D({\mathcal S}^z_{\bm r_\ell})^2  \notag 
	%\\	&&
	+ J \sum_{\ell\in \bigtriangledown}\sum_{j=1}^3\left(
		{\mathcal S}^x_{\bm r_\ell} {\mathcal S}^x_{\bm r_\ell+\bm\delta_j}
		+ {\mathcal S}^y_{\bm r_\ell} {\mathcal S}^y_{\bm r_\ell+\bm\delta_j}
		+ \Delta {\mathcal S}^z_{\bm r_\ell} {\mathcal S}^z_{\bm r_\ell+\bm\delta_j} 
	\right) 
	\notag \\
	&& 
	+ Q \sum_{\ell\in \bigtriangledown}\sum_{i=1}^3 \left({\mathcal S}^y_{\bm r_\ell} {\mathcal S}^y_{{\bm r_\ell}+\bm\delta_i} \cos^2\phi_i + {\mathcal S}^x_{\bm r_\ell} {\mathcal S}^x_{{\bm r_\ell}+\bm\delta_i} \sin^2\phi_i\right) \notag \\
&& + J_{xz} \sum_{\ell\in \bigtriangledown}\sum_{i=1}^3 \left[ \left( {\mathcal S}_{\bm r_\ell}^x \cos\phi_i- {\mathcal S}_{\bm r_\ell}^y \sin\phi_i \right) {\mathcal S}^z_{{\bm r_\ell}+\bm\delta_i} 
%\right. \notag\\
%&& \left. \hspace{1cm}
+ {\mathcal S}^z_{\bm r_\ell} \left( {\mathcal S}_{{\bm r_\ell}+\bm\delta_i}^x \cos\phi_i - {\mathcal S}_{{\bm r_\ell}+\bm\delta_i}^y \sin\phi_i \right) \right], 
	\label{eq:ab-compassXXZ}
	\end{eqnarray}
	where $J=(J_{xx}+J_{yy})/2$, $\Delta=J_{zz}/J$ and $Q=(J_{xx}-J_{yy})/2$ and $\phi_i=2 \pi (i-1)/3$ where $i$ labels the three bonds around each $\bigtriangledown$ 
	shown in Fig. \ref{fig:decorated}. Thus, we see that the second term (proportional to $J$) is simply the XXZ model and the third term (proportional to $Q$) is the honeycomb 120$^\circ$ 
	compass model.\cite{nussinov2015}
\end{widetext}

The exchange couplings and single-spin anisotropy entering model (\ref{eq:ab-compassXXZ}) have been obtained numerically from perturbation theory \cite{merino2017} as well as 
analytically using a canonical transformation.\cite{powell2017} In Fig. \ref{fig:soc} we show the dependence of the model (\ref{eq:ab-compassXXZ}) parameters: $J, Q, \Delta$ and $D$ with $\lambda_{xy}$
for the longitudinal SMOC, $\lambda_z=\lambda_{xy}/2$, appropriate for Mo$_3$S$_7$(dmit)$_3$ crystals. With no SMOC, the exchange
couplings are isotropic: $J \approx 0.04 t_c, Q=0, \Delta =1, J_{xz}=0$, as it should. As SMOC increases, spin exchange anisotropy, {\it i. e.}, $Q \neq 0$, $\Delta \neq 1$  
(or equivalently $J_{xx} \neq J_{yy} \neq J_{zz}$) increases. On the other hand, Fig. \ref{fig:soc} shows how the single-spin anisotropy is rapidly enhanced by SMOC. 

\section{Large-$D$ phase induced by SMOC}

\begin{figure}
 \includegraphics[width=7cm]{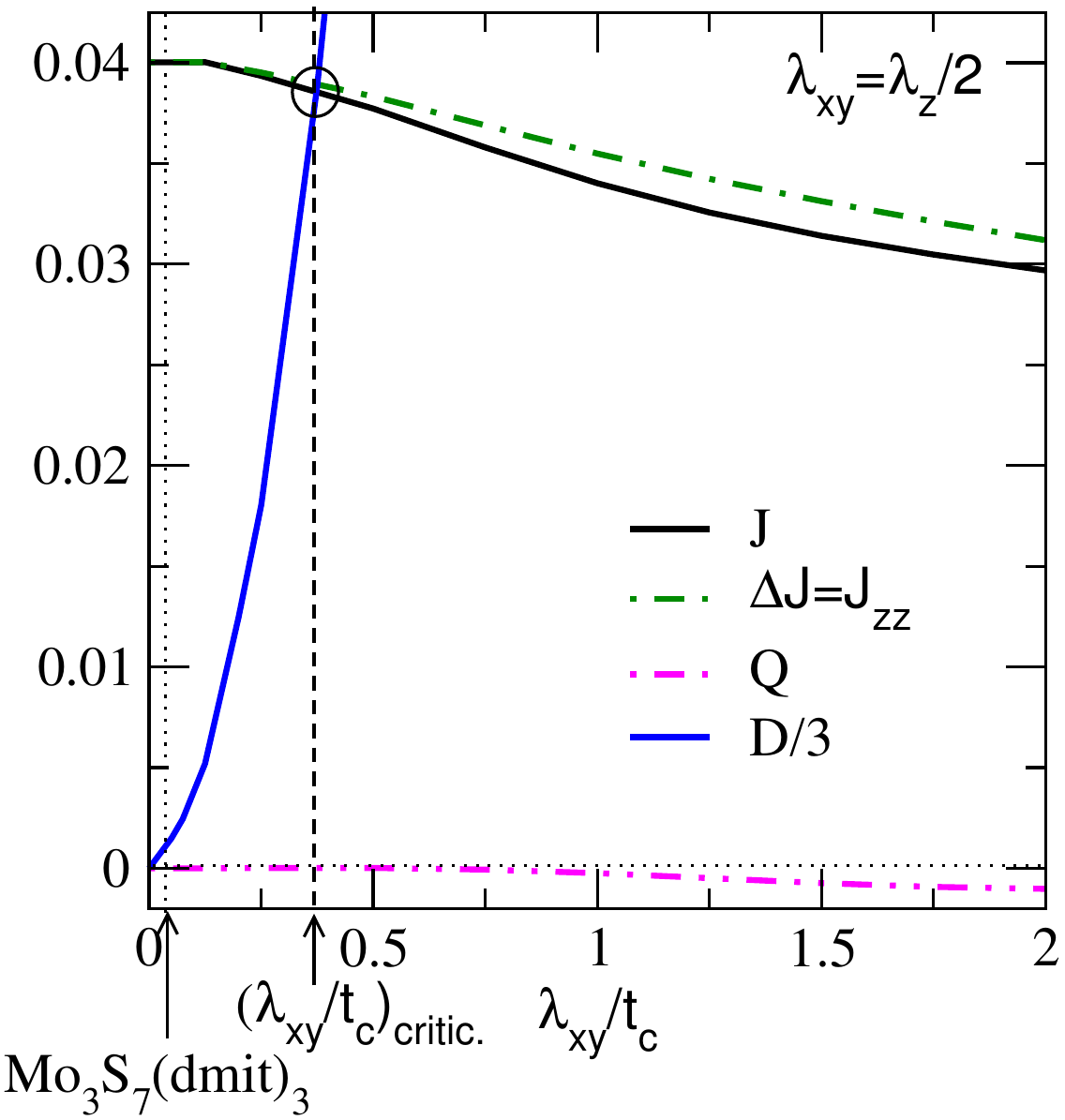}%
 \caption{Spin exchange couplings and single-spin anisotropy in the $a-b$ layers of Mo$_3$S$_7$(dmit)$_3$. 
 The dependence of parameters of model (\ref{eq:ab-compassXXZ}) on SMOC is shown. Longitudinal SMOC is fixed to $\lambda_z=\lambda_{xy}/2$ 
 as appropriate for Mo$_3$S$_7$(dmit)$_3$ and $U=10 t_c$. The
 vertical dashed line marks the critical transverse SMOC, $\lambda_{xy} \approx 0.38 t_c $, at which a transition to 
 a large-$D$ phase occurs. The vertical dotted line marks $\lambda_{xy}/t_c=0.042 t_c$  estimated for Mo$_3$S$_7$(dmit)$_3$ materials.\cite{jacko2017}
 \label{fig:soc}}%
 \end{figure}

One can see from the results of Fig. \ref{fig:soc} that, for sufficiently large SMOC, the single-spin anisotropy overcomes the exchange couplings, $D>J, \Delta J=J_{zz}$. We can expect that in the 
limit, $D \gg J, J_{zz}$, the model is dominated by single-spin anisotropy and the ground state of the model is just the tensor product of ${\mathcal S}_{{\bm r_\ell}}^z=0$
at each lattice site. Hence, it is important to first obtain an estimate of the critical $D_c$ at which a transition to a large-$D$ phase occurs
and the magnitude of SMOC needed to reach such critical $D$. In order to do so, we explore below the magnetic properties of a slightly
simplified version of the full model (\ref{eq:ab-compassXXZ}) in the limit of weak SMOC. This limit is relevant to the $a-b$ planes of Mo$_3$S$_7$(dmit)$_3$,
for which {\it ab initio} calculations\cite{jacko2017} give $\lambda^{\text{Mo}_3\text{S}_7\text{(dmit)}_3}_{xy}=0.042t_c$ with $\lambda_z=\lambda_{xy}/2$. We can 
see from Fig. \ref{fig:soc} that for $\lambda_{xy} \lesssim 0.5$ exchange coupling anisotropies are negligible which justifies using an isotropic Heisenberg model  
for describing the magnetic properties of the $a-b$ planes of the crystal. 

We have recently explored the N\'eel to large-$D$ transition of the isotropic version of model (\ref{eq:ab-compassXXZ}) using a $SU(3)$ description of the spins
which introduces three Schwinger bosons to properly account for the three projections: $S^z_i=0,\pm 1$ of the spin-1 at each lattice site.\cite{merino2018} 
We find a transition from the N\'eel ordered phase to a large-$D$ phase for $D_c \approx 3 J$.
From this condition and the dependence of $D$ with $\lambda_{xy}$  we can obtain a realistic estimate of 
SMOC needed for the large-$D$ phase to occur in isolated $a-b$ layers of Mo$_3$S$_7$(dmit)$_3$. In 
Fig. \ref{fig:soc} we show the critical SMOC $(\lambda_{xy}/t_c)_{critic} \approx 0.38$ at which the $D_c=3J$ condition is satisfied. This critical 
value is about nine times larger than SMOC in Mo$_3$S$_7$(dmit)$_3$ crystals. Hence, we conclude that isolated $a-b$ layers of Mo$_3$S$_7$(dmit)$_3$ should
be in a magnetically ordered N\'eel state.

%\subsubsection{}

\begin{figure}
 \includegraphics[width=7cm]{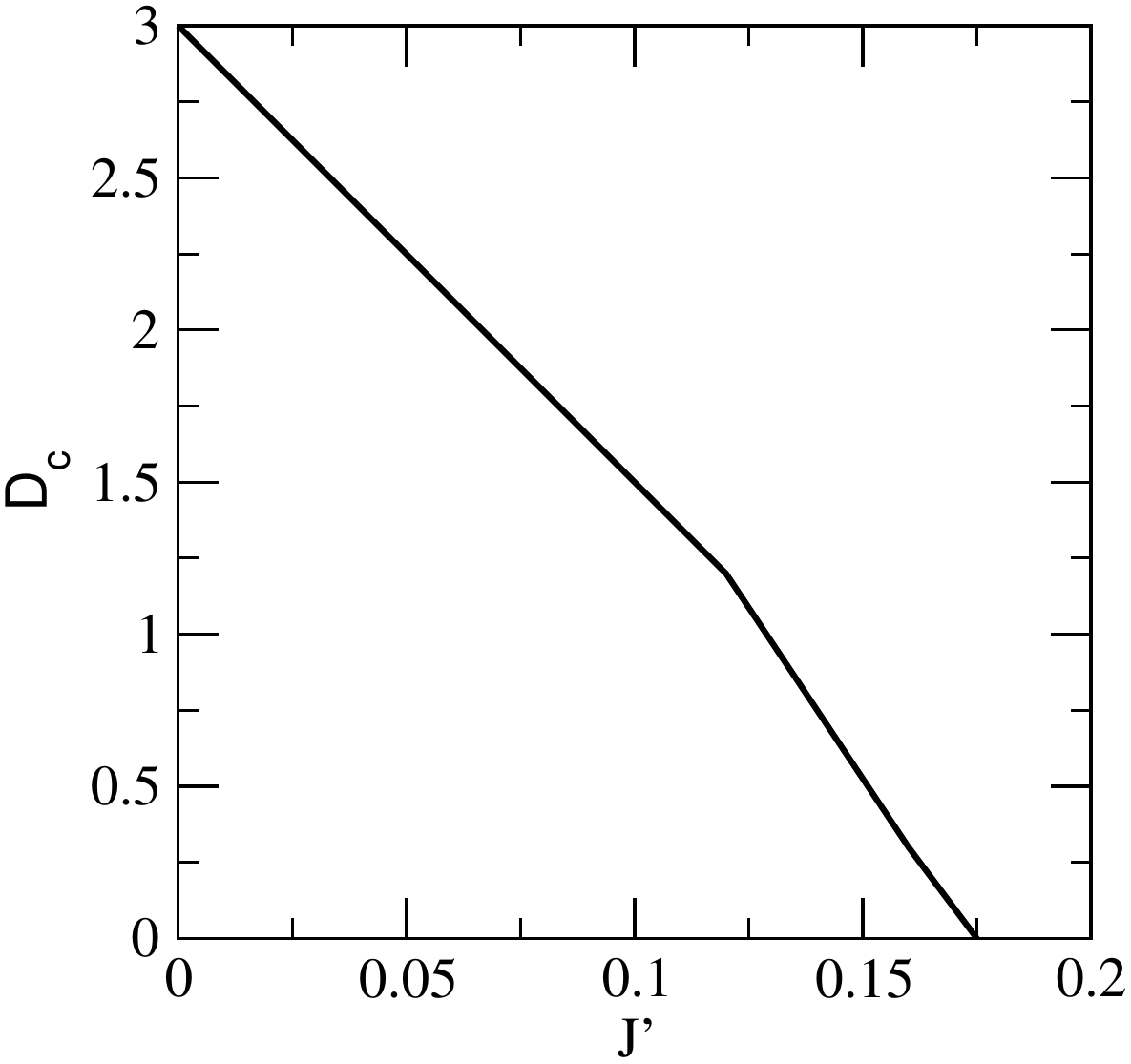}%
 \caption{\label{fig:DcvsJ2} Dependence of the critical single-spin anisotropy with magnetic frustration. The critical $D_c$ 
  at which a transition from a N\'eel phase to a large-$D$ phase occurs from SU(3) SBMFT calculations on model (\ref{eq:j1j2heis}) is shown.
   $J=1$ in this plot.}%
 \end{figure}

\section{Magnetic frustration effects}

Magnetic frustration plays an important role in inducing disordered spin liquid states. 
We analyze the effect of a next-nearest neighbor exchange coupling, $J'$, on the magnetic properties of 
model (\ref{eq:ab-compassXXZ}) in the limit of weak SMOC {\it i. e.} with isotropic couplings. 
Hence, the model we study is a $S=1$ $J$-$J'$ Heisenberg on the honeycomb lattice with single-spin anisotropy:
\begin{eqnarray}
H&=&\sum_\ell D({\mathcal S}^z_{\bm r_\ell})^2 + J \sum_{\langle \ell m \rangle}  {\mathcal S}_{\bm r_\ell}  {\mathcal S}_{\bm r_m}  +J' \sum_{\langle \langle \ell  m \rangle \rangle}  {\mathcal S}_{\bm r_\ell} { \mathcal S}_{\bm r_m}
\notag \\
%&-&{\mu_B B_z \over 2} \sum_ {\ell}  {\mathcal S}^z_{\bm r_{\ell}}, 
\label{eq:j1j2heis}
\end{eqnarray} 
%where we have also added a magnetic field along the $z$-direction, $B_z$, and $\mu_B$ is Bohr's magneton.
We first discuss the phase diagram of this model with no single-spin anisotropy, $D=0$, based on recent SU(2) Schwinger boson mean-field theory (SBMFT)
calculations.\cite{merino2018}  For $J'=0$, the ground state of the model is a N\'eel state, as expected since the lattice is bipartite. A direct transition from N\'eel order to spiral order occurs for $J' \approx 0.24$.    
%For $S=1/2$, SBMFT predicts a quantum spin liquid phase in a small $J'/J$ range between the N\'eel and spiral phases.

% which consists on the tensor product of ${\mathcal S}^z_{{\bm r_\ell}}=0$ at each lattice site. 
%Hence, we expect a transition from a N\'eel to a large-$D$ phase at a critical value of the single-spin anisotropy, $D$. 

We finally discuss the effect of $J'$ on the N\'eel to large-$D$ transition of model (\ref{eq:j1j2heis}) based on SU(3) SBMFT calculations
of model (\ref{eq:j1j2heis}). In Fig. \ref{fig:DcvsJ2} we show the critical $D$ at which a spin 
gap opens up signaling the formation of the paramagnetic large-$D$ phase. We find that the critical $D_c=3J$ obtained for $J'=0$ of the unfrustrated model, 
is strongly suppressed by $J'$. Hence, magnetic frustration favors the large-$D$ phase. The kink around  $J'=0.12 J$ in Fig. \ref{fig:DcvsJ2} is related to the
expected poor performance of the SU(3) SBMFT approach as $D_c$ is suppressed below $J$, $D_c<J$.\cite{merino2018}
%However, increasing $J'/J$ can strongly suppress the critical $D$ as shown in Fig. \ref{fig:DcvsJ2} which means that the large-$D$ phase could 
%be realized even at small SMOC.
% \begin{figure}
 %\includegraphics[width=6cm]{metamagnetJ201T01.eps}%
 %\caption{\label{fig:metamagnet} Metamagnetic transition in the $S=1$, $J-J'$ Heisenberg model on a honeycomb lattice. A transition from a N\'eel ordered
% state to a ferromagnet occurs at a critical field, $B_z$, which is suppressed by the magnetic frustration, $J'$. The temperature is fixed to $T=0.1J$, $D=0$ and $J=1$ in this plot. }%
 %\end{figure}

%We finally discuss the effect of an external field on the N\'eel ordered phase.  In Fig. \ref{fig:metamagnet} we show the dependence of the ferromagnetic moment
%as a function of an external magnetic field applied along the $z$-direction using SBMFT. As the external magnetic field, $B_z$, is increased the ferromagnetic moment increases until a 
%sharp rise to a fully polarized state occurs at a critical magnetic field, $(B_z)_c$.  This kind of transition is commonly known as a metamagnetic transition.\cite{fazekas}
%This critical $(B_z)_c$ is found to be suppressed from a value: $(B_z)_c \sim 2.75 J$ to  $1.5 J$ as $J'$ is increased
%from $J'=0.01 J$ to $0.2 J$. Using {\it ab initio} estimates of $t_c=0.06$ eV  we estimate $J \approx 0.04 t_c=0.0024$ eV in Mo$_3$S$_7$(dmit)$_3$, 
%giving a critical magnetic field in the range  $B_z \approx 114-62$ T for $J' =(0-0.2) J$. Hence, magnetic frustration strongly suppresses the critical field  
%at which the metamagnetic transition occurs. 

\section{Conclusions}

The magnetic properties of the $a-b$ planes of Mo$_3$S$_7$(dmit)$_3$ crystals can be modeled through a $S=1$, XXZ+120$^0$ 
quantum compass model with single-spin anisotropy.\cite{merino2017} For the weak SMOC present in Mo$_3$S$_7$(dmit)$_3$ crystals\cite{jacko2017}, 
quantum compass couplings and single-spin anisotropy play a minor role so that the effective model in this regime is just the isotropic $S=1$ Heisenberg model 
on the honeycomb lattice.  Hence, we predict that isolated honeycomb layers of Mo$_3$S$_7$(dmit)$_3$ are N\'eel ordered. However, we find that 
magnetic frustration induced by a next-nearest neighbor exchange coupling, $J'$, favors the large-$D$ phase. 
%We also predict a metamagnetic transition when 
%a sufficiently strong magnetic field is applied perpendicular to the planes.

% If in two-column mode, this environment will change to single-column format so that long equations can be displayed. 
%\begin{widetext}
%$$\mbox{put long equation here}$$
%\end{widetext}

% Figures should be put into the text as floats. 
% Use the graphics or graphicx packages (distributed with LaTeX2e).
% Put the label that you will use with \ref{} command in the braces of the \label{} command.
%
% \begin{figure}
% \end{figure}

% \begin{table}
% \end{tabular}
% \end{table}
\begin{acknowledgments}
J. M. acknowledges financial support from (MAT2015-66128-R)
MINECO/FEDER, UE, and A.R. from the French National
Research Agency (Contract ORGANIÕSO, ANR-15-CE09-
0017). Work at the University of Queensland was supported by the Australian Research Council (DP160100060).
\end{acknowledgments}


\end{document}